\documentclass[12pt]{article}
\usepackage{amssymb,amsmath,epsfig}
\allowdisplaybreaks

\begin{document}
\title{\textbf{Regular Bardeen Black Hole Solutions in Rastall Theory:
A Gravitational Decoupling Approach}}
\author{M. Sharif$^{1,3}$ \thanks{msharif.math@pu.edu.pk}~
and Malick Sallah$^{1,2}$ \thanks{malick.sallah@utg.edu.gm} \\
$^1$ Department of Mathematics and Statistics, The University of Lahore\\
1-KM Defence Road Lahore-54000, Pakistan.\\
$^2$ Department of Mathematics, The University of The Gambia,\\
Serrekunda, P.O. Box 3530, The Gambia.\\
$^3$ Research Center of Astrophysics and Cosmology, Khazar
University,\\ Baku, AZ1096, 41 Mehseti Street, Azerbaijan.}
\date{}
\maketitle

\begin{abstract}
This research applies the generalized technique of gravitational
decoupling to the Bardeen black hole, producing novel black hole
solutions in the context of Rastall theory. We proceed by
decomposition of the field equations corresponding to an additional
matter source into two sets, for further considerations. The metric
functions of the Bardeen black hole are adopted to specify the first
set. The second one, which is subject to an extra source, is
resolved considering a linear equation of state of matter. Through
the integration of the solutions of these sets, we develop two
expanded models and conduct an in-depth analysis of their distinct
physical characteristics, governed by specific parameters. We
investigate thermodynamic quantities like density, anisotropic
pressure, energy bounds, asymptotic flatness, and thermodynamical
properties like the Hawking temperature, entropy, and specific heat,
etc. Both models are asymptotically flat but violate the energy
bounds. Furthermore, the density, radial pressure, and Hawking
temperature demonstrate consistent and acceptable behavior.
Ultimately, the thermodynamic stability is affirmed through the
analysis of specific heat and the Hessian matrix.
\end{abstract}

\section{Introduction}

Rastall \cite{5} first questioned the principle of energy-momentum
conservation in curved spacetime, proposing a non-minimal
interaction between matter and geometric fields. Evidence supporting
this theory emerged from observations of particle production during
cosmic evolution \cite{6}-\cite{8}. However, the Rastall theory has
faced criticism \cite{1a,1aa}, particularly regarding the lack of
conservation of energy-momentum tensor, a claim contested by other
authors \cite{1b}. However, this perceived violation can be seen as
a consequence of spacetime curvature or even the net creation of
energy in certain systems. Another common criticism is the absence
of a Lagrangian formulation for the theory, despite its success in
producing acceptable results in both cosmology and astrophysics.
Attempts to derive a suitable Lagrangian have been unsuccessful so
far, raising doubts about its feasibility. Despite these challenges,
the advantages of the Rastall theory are quite notable, with various
theoretical and observational studies appearing in recent research
\cite{2}-\cite{4a}. In more recent studies, the role of the Rastall
parameter in constructing novel stellar solutions within spherical
symmetry \cite{5a} as well as in various models involving complexity
and isotropization \cite{5aa,5aaa} has been investigated.

Black holes are some of the most remarkable entities in the cosmos.
Their direct observation via gravitational wave detection
\cite{14,15} and the imaging of black hole shadows \cite{16} has
elevated them from theoretical constructs in general relativity (GR)
to tangible astrophysical phenomena with clearly defined attributes.
According to GR, black holes are characterized by just three
essential properties: mass $(M)$, charge $(Q)$, and angular momentum
$(J)$ \cite{17}. The no-hair theorem suggests that no additional
charge should exist in these solutions \cite{18}. Nevertheless, it
has been suggested that black holes may also harbor other charges
linked to internal gauge symmetries and fields, and they might even
exhibit (soft) quantum hair \cite{18a}. Many solutions of the GR
field equations reveal future singularities \cite{19} and past
singularities \cite{20}-\cite{23}. This becomes complicated with the
presence of black holes, as the singularity is generally hidden
behind the event horizon \cite{24}. One of the important turns in
the development of black hole physics is associated with the work of
Stephen Hawking \cite{25,26} who showed that black hole horizon
emits radiation. As a result of this, black holes became important
as they provided a playground to test the theories of gravitation in
a quite elaborate way.

Solutions of the Einstein field equations that are devoid of a
singularity are termed as regular black holes. If the strong energy
condition holds, then the collapse of matter clouds lead to the
formation of singularities, as specified by the singularity theorem
of Penrose \cite{19}. This theorem, though, can be bypassed if one
considers the distortion of the black hole center by violating the
strong energy condition. If $lim_{r\to 0}\mathcal{K}=\pm\infty$,
where $\mathcal{K}=\mathcal{R}_{abcd}\mathcal{R}^{abcd}$ is the
Kretschmann scalar, then the presence of a singularity is affirmed.
The Bardeen black hole solution \cite{27}, which is is supported by
nonlinear electrodynamics \cite{28}, is the first known regular
black hole. The existence of singularities suggests that GR might
not be completely applicable in regions of extremely high mass
concentration. This has led to increased interest in black hole
models that do not include singular centers \cite{29}-\cite{34}.

The study of regular black holes within modified gravity frameworks
provides valuable insights into the nature of singularity resolution
and alternative gravitational dynamics. In this context, extending
the Bardeen black hole to Rastall gravity allows us to explore how
the non-conservative energy-momentum framework influences regular
black hole structures. Moreover, the gravitational decoupling method
serves as a powerful tool to systematically introduce anisotropic
modifications while preserving key physical properties. By employing
this approach, we construct and analyze new black hole solutions,
examining their energy conditions, thermodynamic stability, and the
role of the Rastall parameter in shaping their characteristics. This
investigation not only broadens our understanding of regular black
holes in alternative theories but also highlights the impact of
energy-momentum exchange in gravitational systems.

A key aspect of fundamental physics is the study of black hole
thermodynamics. Indeed, concepts such as specific heat and entropy
form an important bridge between GR, quantum theory, and stochastic
mechanics. Black holes and their evaporation have been examined by
scientists in order to better understand spacetime, the quantum
fields in strong gravity, and the laws of the universe evolution.
Our understanding of black holes as well as the fundamental
principles of physics in extreme environments is expanded by this
integrated approach. The relationship between the surface area and
entropy of black holes was initially studied by Bekenstein
\cite{35}, while Hawking \cite{25} showed that with surface gravity,
$k$, black holes can be thought of as thermal sources when their
temperature is $\frac{k}{2\pi}$. However, the Bekenstein-Hawking
radiation raises the question of the loss of information owing to
heat loss. To circumvent this problem, soft hair was proposed by
Hawking et al. \cite{18a}. Due to the contributions of Bekenstein
and Hawking to black hole thermodynamics, the radiation from black
holes became of great concern among researchers. A recent trend in
the research of black holes is their thermodynamical analysis within
the Rastall gravity theory \cite{36}-\cite{40}.

To develop innovative models of relativistic bodies exhibiting
diverse characteristics, the gravitational decoupling approach was
proposed \cite{41}. This method has been proven effective in
theoretically constructing potential star distributions under
various conditions \cite{42}-\cite{46}. Numerous researchers
\cite{47,47a} have utilized this approach to describe the interior
of self-gravitating structures having an anisotropic distribution of
hybrid fluids. Due to the high nonlinearity of the field equations,
there are few physically plausible analytical solutions available
unless certain constraints are applied. The gravitational decoupling
technique involves splitting the matter source into two parts; the
initial sector is chosen in order to offer a well known solution of
GR, while the later one corresponds to an extra source. The
gravitational decoupling method consists of two fundamental
approaches known as minimal and extended geometric deformations (MGD
and EGD, respectively). The basic difference between these
approaches is that the MGD only alters the $g_{rr}$ component of the
metric, whereas in EGD, both the $g_{rr}$ and $g_{tt}$ parts are
changed. Moreover, MGD is limited to cases where the interaction
between the sources is strictly gravitational and hence cannot be
applied when there is non-gravitational interaction (such as energy
exchange) between the sources.

Ovalle \cite{48} introduced the concept of EGD by modifying both
metric components, \( g_{rr} \) and \( g_{tt} \). Since then, this
approach has been widely employed to construct anisotropic spherical
solutions across various alternative gravity theories. In the
context of $(2+1)$-dimensional spacetimes, the EGD technique has
been applied to extend the BTZ model \cite{49}. Sharif and Mughani
\cite{50} utilized this framework to generate extensions of certain
isotropic solutions, while generalizations of the Tolman IV and
Krori-Barua models have been explored within the Brans-Dicke theory
\cite{51}. Ovalle and his collaborators \cite{52} further leveraged
this method to obtain hairy black hole solutions by extending the
vacuum Schwarzschild metric. Additionally, the influence of an
electric field on decoupled solutions within the Brans-Dicke
framework was analyzed in \cite{53}. The same group of researchers
\cite{54} extended this approach to derive a generalized
Schwarzschild black hole solution within the Brans-Dicke theory.
Sharif and Naseer \cite{55} also applied this methodology to
construct a range of extended models in different modified gravity
theories. Lastly, in our own work, we have employed this scheme to
obtain a generalized Schwarzschild black hole solution \cite{56} as
well as an extension of the Tolman IV ansatz within an electric
field \cite{56aa} in the framework of Rastall gravity.

Recent advances in black hole thermodynamics and deformation
techniques offer valuable context for the present study. In
\cite{56d}, researchers examined horizon thermodynamics and
established links between geometric quantities and thermodynamic
laws. Ali et al. \cite{56dd} analyzed restricted phase--space
thermodynamics for rotating AdS black holes, while Pourhassan et al.
\cite{56ddd} and Soroushfar et al. \cite{56e} investigated quantum
and holographic corrections to black hole thermodynamics. Hazarika
and Phukon \cite{56ee} further explored topological aspects of such
corrections within extended phase spaces. Collectively, these works
support the broader relevance of exploring modified gravity
frameworks such as the present Rastall-EGD scenario where geometric
deformation and non--conservative effects enrich the thermodynamic
behavior of regular black holes.

This study employs the EGD method to derive new regular black holes
within Rastall theory, yielding two new solutions. These novel
solutions are examined and analyzed against previous research. We
organize this work in the following sequence: The Rastall field
equations with a more general matter source are presented in section
\textbf{2}. In the next section, \textbf{3}, the same field
equations are studied using the EGD approach. Section \textbf{4}
produces and studies the new models in detail, emphasizing on the
deformed metric potentials, their asymptotic flatness, as well as
energy conditions. In section \textbf{5}, the new solutions are
subjected to a thermodynamic analysis. A summary of the results with
concluding remarks are provided in section \textbf{6}.

\section{Field Equations}

Rastall's field equations differ from Einstein's field equations due
to the presence of the Rastall parameter, $\zeta$. This parameter
establishes a connection between the covariant derivative of the
Rastall stress-energy tensor and that of the curvature scalar,
$\mathcal{R}$. Consequently, the Rastall field equations are
expressed as follows
\begin{equation}\label{1}
G_{\eta\xi}+\frac{\zeta}{4}\mathcal{R}g_{\eta\xi}=\kappa
T^R_{\eta\xi},
\end{equation}
and are consistent with the relation
\begin{equation}\label{2}
\nabla^\xi T^R_{\eta\xi}=\frac{\zeta}{4}
g_{\eta\xi}\nabla^\xi\mathcal{R}.
\end{equation}
Here, the Einstein tensor is denoted by $G_{\eta\xi}$, while
$\kappa$ and $g_{\eta\xi}$ are the coupling constant and metric
tensor, respectively. Contracting the field equations \eqref{1} and
using the resulting expression for the curvature scalar,
$\mathcal{R}$, we obtain the equivalent form
\begin{equation}\label{3}
G_{\eta\xi}=\kappa\bigg(T^R_{\eta\xi}-\frac{\zeta}{4(\zeta-1)}T^R
g_{\eta\xi}\bigg).
\end{equation}
This equation can be expressed as
\begin{equation}\label{4}
G_{\eta\xi}=\kappa T_{\eta\xi},
\end{equation}
by defining
\begin{equation}\label{5}
T_{\eta\xi}=T^R_{\eta\xi}-\frac{\zeta}{4(\zeta-1)}T^R g_{\eta\xi}.
\end{equation}

It is important to clarify that the introduction of the effective
energy-momentum tensor in Eq.\eqref{5} above is not intended to
imply formal equivalence between Rastall gravity and GR, as argued
in \cite{1a}, but rather to express the field equations in a GR-like
form that facilitates analysis. In Rastall gravity, the divergence
of the energy-momentum tensor is not zero (Eq.\eqref{2}), which
introduces a non-conservative aspect to the gravitational dynamics.
This distinguishes it physically from GR, especially in the context
of matter interactions and thermodynamics, as several studies have
shown deviations in astrophysical predictions \cite{36}-\cite{40}.
Our use of the effective form is strictly for convenience in
expressing the equations and does not suggest a reinterpretation of
the theory's foundational principles.

We identify $T_{\eta\xi}$ as the anisotropic fluid energy-momentum
tensor, given by
\begin{equation}\label{6}
T_{\eta\xi}=(\rho+P_t)V_\eta V_\xi-P_tg_{\eta\xi}+(P_r-P_t)Y_\eta
Y_\xi.
\end{equation}
From Eq.\eqref{6} above, $\rho,P_r,P_t,$
$V_\eta=(\sqrt{g_{00}},0,0,0)$ and $Y_\eta=(0,-\sqrt{-g_{11}},0,0)$
denote the density, radial pressure, transverse pressure, 4-vector
and 4-velocity, respectively, and satisfy
\begin{equation}\nonumber
V^\eta Y_\eta=0,\quad V^\eta V_\eta=1,\quad Y^\eta Y_\eta=-1.
\end{equation}
By contracting Eq.\eqref{5}, we deduce that
\begin{equation}\label{7}
(1-\zeta)T=T^R,
\end{equation}
with which we can write
\begin{equation}\label{8}
T^R_{\eta\xi}=T_{\eta\xi}-\frac{\zeta}{4}Tg_{\eta\xi}.
\end{equation}
The addition of a extra source to a seed source forms the basis on
which the gravitational decoupling scheme is founded. The field
equations are thus enhanced as follows
\begin{equation}\label{9}
G_{\eta\xi}+\frac{\zeta}{4}\mathcal{R}g_{\eta\xi}=\kappa
\hat{T}_{\eta\xi},
\end{equation}
where
\begin{equation}\label{10}
\hat{T}_{\eta\xi}=T^R_{\eta\xi}+\psi\Omega_{\eta\xi}.
\end{equation}
The equation above demonstrates that the total energy-momentum
tensor consists of a principal component, $T^R_{\eta\xi}$, which is
gravitationally connected to an additional matter source,
$\Omega_{\eta\xi}$, via the decoupling parameter $\psi$. This extra
source may involve new scalar, vector, and tensor fields,
contributing to the anisotropy observed in the fluid.

The following metric describes the spacetime geometry
\begin{equation}\label{12}
ds^2=e^{\chi_1(r)}dt^2-e^{\chi_2(r)}dr^2-r^2(d\theta^2+\sin^2\theta
d\phi^2).
\end{equation}
With this metric, the field equations \eqref{9} become
\begin{align}\nonumber
\kappa\left[\rho-\frac{\zeta}{4}(\rho-P_r-2P_t)+\psi\Omega^0_0
\right]&=\frac{1}{r^2}+e^{-\chi_2}\bigg(\frac{\chi_2^\prime}{r}
-\frac{1}{r^2}\bigg)+\frac{\zeta e^{-\chi_2}}{4}
\bigg(\chi_1^{\prime\prime}+\frac{\chi_1^\prime(\chi_1^\prime
-\chi_2^\prime)}{2}\bigg)\\\label{13} &+\frac{\zeta
e^{-\chi_2}}{4}\bigg(\frac{2(\chi_1^\prime -\chi_2^\prime)}{r}
+\frac{2}{r^2}\bigg)-\frac{\zeta}{2r^2},
\end{align}
\begin{align}\nonumber
\kappa\left[P_r+\frac{\zeta}{4}(\rho-P_r-2P_t)-\psi\Omega^1_1\right]&=
-\frac{1}{r^2}+e^{-\chi_2}\bigg(\frac{\chi_1^\prime}{r}+\frac{1}{r^2}\bigg)
-\frac{\zeta e^{-\chi_2}}{4}\bigg(\chi_1^{\prime\prime}
+\frac{\chi_1^\prime(\chi_1^\prime-\chi_2^\prime)}{2}\bigg)\\\label{14}
&-\frac{\zeta e^{-\chi_2}}{4}\bigg(\frac{2(\chi_1^\prime
-\chi_2^\prime)}{r} +\frac{2}{r^2}\bigg)+\frac{\zeta}{2r^2},
\end{align}
\begin{align}\nonumber
\kappa\left[P_t+\frac{\zeta}{4}(\rho-P_r-2P_t)-\psi\Omega^1_1\right]&=
e^{-\chi_2}\bigg(\frac{\chi_1^{\prime\prime}}{2}
+\frac{\chi_1^{\prime^2}}{4} -\frac{\chi_1^\prime
\chi_2^\prime}{4}+\frac{\chi_1^\prime}{2r}
-\frac{\chi_2^\prime}{2r}\bigg)+\frac{\zeta}{2r^2}\\\label{15}
&-\frac{\zeta e^{-\chi_2}}{4}\bigg(\chi_1^{\prime\prime}
+\frac{\chi_1^\prime(\chi_1^\prime-\chi_2^\prime)}{2}
+\frac{2(\chi_1^\prime-\chi_2^\prime)}{r}+\frac{2}{r^2}\bigg).
\end{align}
This system constitutes eight unknowns
$(\chi_1,\chi_2,P_r,\rho,P_t,\Omega^0_0,\Omega_1^1,\Omega^2_2)$ in
three differential equations. Additionally, the prime notation
denotes the radial derivative. From this system, we define the
matter variables
\begin{equation}\label{17}
\tilde{\rho}=\rho+\psi\Omega^0_0,\quad
\tilde{P_r}=P_r-\psi\Omega^1_1,\quad \tilde{P_t}=P_t-\psi\Omega^2_2,
\end{equation}
which induce an anisotropy given by
\begin{equation}\label{18}
\hat{\Pi}=\tilde{P_t}-\tilde{P_r}=(P_t-P_r)+\psi(\Omega_1^1-\Omega^2_2).
\end{equation}

In the subsequent analysis, we employ the EGD approach to decouple
the field equations \eqref{13}-\eqref{15}. With this approach, the
field equations are decomposed into two different groups. The first
group relates to the primary source and is described by the metric
of the Bardeen black hole \cite{27}. The second group takes into
account the additional source which will be addressed through
suitable restrictions.

\section{Extended Geometric Deformation}

The complexity in the field equations increases upon introducing an
added source to the initial fluid, leading to the emergence of new
parameters. It is necessary to impose a restriction on the degrees
of freedom in order to be able to arrive at a solution by adopting a
particular method or a particular set of conditions. We thus apply
the well known technique of gravitational decoupling, through which
the field equations are solved. An interesting aspect of this method
is its ability to transform the metric functions into a new frame of
reference without distorting the spacetime geometry. Specifically,
we use the EGD scheme wherein both $g_{rr}$ and $g_{tt}$ metric
components are transformed. Subsequently, we analyze a known
anisotropic fluid solution to the field equations, defined by the
metric
\begin{equation}\label{19}
ds^2=e^{\chi_3(r)}dt^2-\frac{1}{\chi_4(r)}dr^2-r^2(d\theta^2+\sin^2\theta
d\phi^2),
\end{equation}
where
\begin{equation}\label{20}
\chi_4(r)=1-\frac{2m(r)}{r},
\end{equation}
with $m$ as the Misner-Sharp mass. The geometric deformation are
encoded in the transformations
\begin{equation}\label{21}
\chi_1(r)=\chi_3(r)+\psi h^\ast(r),\quad
e^{-\chi_2(r)}=\chi_4(r)+\psi g^\ast(r),
\end{equation}
where $g^\ast(r)$ and $h^\ast(r)$ deform $g_{rr}$ and $g_{tt}$,
respectively. Through these transformations, the field equations are
decoupled into two sets, the first of which is
\begin{align}\nonumber
\kappa\left[\rho-\frac{\zeta}{4}(\rho-P_r-2P_t)\right]&
=\chi_4\bigg(\frac{\zeta\chi_3^{\prime\prime}}{4}
-\frac{1}{r^2}+\frac{\zeta\chi_3^{\prime^2}}{8}
+\frac{\zeta\chi_3^\prime}{2r} +\frac{\zeta}{2r^2}\bigg)\\\label{22}
&+\chi_4^\prime\bigg(\frac{\zeta}{2r}
+\frac{\zeta\chi_3^\prime}{8}-\frac{1}{r}\bigg)
-\frac{\zeta}{2r^2}+\frac{1}{r^2},
\end{align}
\begin{align}\nonumber
\kappa\left[P_r+\frac{\zeta}{4}(\rho-P_r-2P_t)\right]&
=\chi_4\bigg(\frac{\chi_3^\prime}{r}
-\frac{\zeta\chi_3^{\prime\prime}}{4}+\frac{1}{r^2}
-\frac{\zeta\chi_3^{\prime^2}}{8}-\frac{\zeta\chi_3^\prime}{2r}
-\frac{\zeta}{2r^2}\bigg)\\\label{23}
&-\chi_4^\prime\bigg(\frac{\zeta\chi_3^\prime}{8}
+\frac{\zeta}{2r}\bigg)+\frac{\zeta}{2r^2}-\frac{1}{r^2},
\end{align}
\begin{align}\nonumber
\kappa\left[P_t+\frac{\zeta}{4}(\rho-P_r-2P_t)\right]&
=\chi_4\bigg(\frac{\chi_3^{\prime\prime}}{2}
+\frac{\chi_3^{\prime^2}}{4}+\frac{\chi_3^\prime}{2r}
-\frac{\zeta\chi_3^{\prime\prime}}{4}
-\frac{\zeta\chi_3^{\prime^2}}{8}-\frac{\zeta\chi_3^\prime}{2r}
-\frac{\zeta}{2r^2}\bigg)\\\label{24}&+\chi_4^\prime\bigg(\frac{\chi_3^\prime}{4}
+\frac{1}{2r}-\frac{\zeta\chi_3^\prime}{8}-\frac{\zeta}{2r}\bigg)+\frac{\zeta}{2r^2}.
\end{align}

From this system, we observe five parameters
$\big(\chi_3,P_r,\rho,P_t,\chi_4\big)$, in three equations. To
complete this system, it is sufficient to implement two constraints.
For this purpose, we will use the metric coefficients of the Bardeen
black hole \cite{27}. Although the Bardeen black hole is originally
derived in GR, its metric structure can still be used as a seed
solution in Rastall gravity, provided that the associated
energy-momentum tensor is appropriately adjusted to satisfy the
modified field equations. This approach has been successfully
employed in various extensions of known GR solutions to alternative
gravity frameworks \cite{54}, \cite{56a}-\cite{56c}. We emphasize
that the Bardeen metric used as a seed solution is introduced as a
geometric ansatz, and its associated energy-momentum tensor is
recomputed within the Rastall framework. Thus, it satisfies its own
decoupled sector of the field equations derived from Rastall gravity
(Eqs.(20)-(22)). This approach preserves the core principle of
gravitational decoupling, namely, that the seed metric must fulfill
the undeformed field equations in order to justify the separation of
sources \cite{48}. Any extensions or deformations added via the
extra source term are constructed to satisfy the remaining sector of
the modified field equations.

The second set which entails the additional matter source,
$\Omega_{\eta\xi}$, is described by the system
\begin{align}\nonumber
\kappa\Omega^0_0&=\frac{\zeta}{4}\bigg[g^\ast\bigg(
\chi_3^{\prime\prime}+\frac{\chi_3^{\prime^2}}{2}
+\frac{2\chi_3^\prime}{r}\bigg)+g^{\ast^\prime}\bigg(
\frac{\chi_3^\prime}{2}+\frac{2}{r}\bigg)+\nu_4
h^{\ast^{\prime\prime}}+\chi_4\chi_3^\prime
h^{\ast^\prime}+\frac{\chi_4\psi h^{\ast^{\prime^2}}}{2}\\\label{26}
&+\frac{\chi_4^\prime h^{\ast^\prime}}{2}+\frac{2\chi_4
h^{\ast^\prime}}{r}\bigg]-\frac{g^{\ast^\prime}}{r}-\frac{g^\ast}{r^2},
\end{align}
\begin{align}\nonumber
\kappa\Omega^1_1&=\frac{\zeta}{4}\bigg[g^\ast\bigg(
\chi_3^{\prime\prime}+\frac{\chi_3^{\prime^2}}{2}
+\frac{2\chi_3^\prime}{r}\bigg)+g^{\ast^\prime}\bigg(
\frac{\chi_3^\prime}{2}+\frac{2}{r}\bigg)+\chi_4
h^{\ast^{\prime\prime}}+\chi_4\chi_3^\prime
h^{\ast^\prime}+\frac{\chi_4\psi h^{\ast^{\prime^2}}}{2}\\\label{27}
&+\frac{\chi_4^\prime h^{\ast^\prime}}{2}+\frac{2\chi_4
h^{\ast^\prime}}{r}\bigg]-g^\ast\bigg(\frac{\chi_3^\prime}{r}
+\frac{1}{r^2}\bigg)-\frac{\chi_4 h^{\ast^\prime}}{r},
\end{align}
\begin{align}\nonumber
\kappa\Omega^2_2&=\frac{\zeta}{4}\bigg[g^\ast\bigg(
\chi_3^{\prime\prime}+\frac{\chi_3^{\prime^2}}{2}
+\frac{2\chi_3^\prime}{r}\bigg)+g^{\ast^\prime}\bigg(
\frac{\chi_3^\prime}{2}+\frac{2}{r}\bigg)+\chi_4
h^{\ast^{\prime\prime}}+\chi_4\chi_3^\prime
h^{\ast^\prime}+\frac{\chi_4\psi
h^{\ast^{\prime^2}}}{2}\\\nonumber&+\frac{\chi_4^\prime
h^{\ast^\prime}}{2}+\frac{2\chi_4 h^{\ast^\prime}}{r}\bigg]
-g^\ast\bigg(\frac{\chi_3^{\prime\prime}}{2}
+\frac{\chi_3^{\prime^2}}{4}+\frac{\chi_3^\prime}{2r}\bigg)
-g^{\ast^\prime}\bigg(\frac{\chi_3^\prime}{4}+\frac{1}{2r}\bigg)
-\frac{\chi_4^\prime h^{\ast^\prime}}{4}\\\label{28}&-\chi_4\bigg(
\frac{h^{\ast^{\prime\prime}}}{2}+\frac{\psi h^{\ast^{\prime^2}}}{4}
+\frac{\chi_4^\prime
h^{\ast^\prime}}{2}+\frac{h^{\ast^\prime}}{2r}\bigg).
\end{align}
To solve this system of equations, two constraints are imposed to
regulate five unknowns distributed across three equations. The first
constraint is enforced on the metric potentials, while the second
pertains to an additional source characterized by a linear equation
of state (EoS). A solution to the field equations is subsequently
derived using the superposition principle, wherein a linear
combination of solutions from the preceding systems is constructed.
This combination is meticulously chosen in accordance with the
effective parameters defined in Eq.\eqref{17}.

\section{Extended Bardeen Black Hole}

A key consideration in extending the Bardeen black hole to Rastall
gravity is ensuring consistency with the modified field equations.
While the original Bardeen solution was obtained in GR with
nonlinear electrodynamics as a source, its metric structure can
still be employed within Rastall gravity, provided that the
energy-momentum tensor is properly modified to satisfy the theory's
equations. Rastall gravity does not alter the geometric sector of
the field equations but rather introduces a non-conservative
energy-momentum framework. This implies that known GR solutions can
still serve as valid models in Rastall gravity if their associated
matter content is appropriately reinterpreted. In this work, we
ensure this consistency through the gravitational decoupling method,
which allows us to introduce an additional anisotropic source that
balances the Rastall-modified field equations. This approach not
only extends the Bardeen solution within a new theoretical framework
but also provides insights into the influence of the Rastall
parameter on the thermodynamic and stability properties of regular
black holes.

The metric for the Bardeen black hole \cite{27} is given by
\begin{equation}\label{30}
ds^2=\bigg(1-\frac{2Mr^2}{(r^2+e^2)^{\frac{3}{2}}}\bigg)dt^2-
\bigg(1-\frac{2Mr^2}{(r^2+e^2)^{\frac{3}{2}}}\bigg)^{-1}dr^2
-r^2(d\theta^2+\sin^2\theta d\phi^2).
\end{equation}
Here $e$ is the magnetic charge and $M$ is the mass of the black
hole. This metric features both a Killing horizon and a causal
horizon located at the surface defined by $r_H$ and $r_h$
respectively. In order to define these horizons, there are
conditions that need to be given, such as $e^{\chi_1}=0$ and
$e^{-\chi_2}=0$ respectively \cite{58}. We thus obtain
\begin{equation}\label{31}
r_H=\sqrt{\tilde{B_1}e^2+\frac{\tilde{B_2}+\tilde{B_3}}{3}+\frac{4M^2}{3}},
\end{equation}
where
\begin{align}\nonumber
\tilde{B_1}&=-\frac{2^{8/3} M^2}{\sqrt[3]{27 e^4 M^2-72 e^2 M^4+3
\sqrt{81 e^8 M^4-48 e^6 M^6}+32 M^6}}-1,\\\nonumber
\\\nonumber \tilde{B_2}&=\frac{
2^{11/3}M^4}{\sqrt[3]{27 e^4 M^2-72 e^2 M^4+3 \sqrt{81 e^8 M^4-48
e^6 M^6}+32 M^6}},\\\nonumber
\\\nonumber \tilde{B_3}&=\left(54 e^4 M^2+64 M^6-144 e^2 M^4+6
\sqrt{81 e^8 M^4-48 e^6 M^6}\right)^{\frac{1}{3}}.
\end{align}
By deforming the metric \eqref{30}, we derive generalizations of the
regular Bardeen black hole solution. The metric for the deformed
Bardeen black hole reads
\begin{equation}\label{32}
ds^2=\bigg(1-\frac{2Mr^2}{(r^2+e^2)^{\frac{3}{2}}}\bigg)e^{\psi
h^\ast(r)}dt^2-\frac{dr^2}{\bigg(1-\frac{2Mr^2}{(r^2+e^2)^{\frac{3}{2}}}
+\psi g^\ast(r)\bigg)} -r^2(d\theta^2+\sin^2\theta d\phi^2),
\end{equation}
where $h^\ast(r)$ and $g^\ast(r)$ are obtained from the system
\eqref{26}-\eqref{28}. The deformed metric Eq.\eqref{33} describes
the extended models and is characterized by the matter variables
\begin{align}\nonumber
\tilde{\rho}&=\frac{3 e^2 M \left(4 e^2 (\zeta -1)-(\zeta +4)
r^2\right)}{16 \pi  (\zeta -1)
\left(e^2+r^2\right)^{7/2}}+\psi\Omega_0^0,\\\nonumber\tilde{P_r}&=\frac{3
e^2 M \left((\zeta +4) r^2-4 e^2 (\zeta-1)\right)}{16 \pi (\zeta-1)
\left(e^2+r^2\right)^{7/2}}-\psi\Omega_1^1,
\\\label{33}\tilde{P_t}&=-\frac{3 e^2 M \left(4 e^2 (\zeta-1)+(6-11\zeta)
r^2\right)}{16\pi(\zeta-1)\left(e^2+r^2\right)^{7/2}}-\psi\Omega_2^2.
\end{align}

The simultaneity of the horizons represents a sufficient condition
for Eq.\eqref{32} to depict a well-defined black hole. Owing to this
coincidence we have $e^\chi_1=e^{-\chi_2}$, which implies that
\begin{equation}\label{34}
\chi_1=-\chi_2.
\end{equation}
The relation $\chi_1=-\chi_2$ ensures the preservation of a
Schwarzschild-type metric structure, where $g_{tt}=g_{rr}^{-1}$.
This ensures that the event horizon determined from $g_{tt}(r_H)=0$
coincides with the causal horizon where $g_{rr}^{-1}(r_H)$, thus
yielding a single, well-defined horizon and surface gravity. If this
condition were relaxed, the metric components would in general
vanish at different radii, leading to multiple horizons or
non-static configurations whose thermodynamics require a separate
treatment. Although such cases may be of interest for future
exploration, we focus here on the coincident-horizon scenario to
maintain direct correspondence with the standard Bardeen and
Rastall-Bardeen geometries. It is worthy of mentioning that this
condition, as has being exploited here, was also used in different
studies on black holes within GR \cite{52,58} as well as modified
theories \cite{54,58a,58ab}.

By incorporating the constraint $\chi_1=-\chi_2$ into the
transformation equations \eqref{21}, we derive the fundamental
relationship that governs the deformation functions, expressed as
\begin{equation}\label{35}
g^\ast(r)=\frac{\left(e^{\psi h^\ast(r)}-1\right)
\bigg(1-\frac{2Mr^2}{(r^2+e^2)^{\frac{3}{2}}}\bigg)}{\psi}.
\end{equation}
The equation above acts as a closure condition that links the radial
and temporal deformation functions within the EGD framework. This
correlation is physically motivated as it preserves the
Schwarzschild-like symmetry $g_{tt}=g_{rr}^{-1}$, ensures regular
behavior of the geometry at the origin, and guarantees asymptotic
flatness. The relation also restricts the number of free functions
so that the modified field equations become determinate. Other
possible couplings between $g^{\ast}(r)$ and $h^{\ast}(r)$ could
produce distinct horizon or anisotropy profiles, but we focus on the
current relation as it yields regular, single-horizon solutions
consistent with the Bardeen framework.

The linear EoS \cite{58}
\begin{equation}\label{36}
\Omega_0^0+\mu\Omega_1^1=\nu\Omega^2_2,
\end{equation}
where $\mu,\nu\in\mathbb{R}$, is employed as the second constraint
needed in obtaining $g^\ast$ and $h^\ast$.

Consequently, two extended solutions with regular features are
deduced in the subsequent sections, making use of the two cases of
EoS mentioned earlier.

\subsection{Model I: Conformally Symmetric $\Omega_{\eta\xi}$}

If the stress-energy tensor of the matter source $\Omega_{\eta\xi}$
is traceless, then this source is said to be conformally symmetric.
Since $\Omega_2^2=\Omega^3_3$, then $\Omega_{\eta\xi}$ is traceless
if
\begin{equation}\label{37}
\Omega_0^0+\Omega^1_1+2\Omega^2_2=0.
\end{equation}
Using \eqref{26}-\eqref{28}, Eq.\eqref{37} becomes
\begin{align}\nonumber
&-g^\ast\bigg(\frac{\chi_3^\prime}{r}+\frac{2}{r^2}\bigg)
-\frac{g^{\ast^\prime}}{r}-\frac{\chi_4 h^{\ast^\prime}}{r}
+\zeta\bigg[g^\ast\bigg(\chi_3^{\prime\prime}
+\frac{\chi_3^{\prime^2}}{2}+\frac{2\chi_3^\prime}{r}\bigg)
+g^{\ast^\prime}\bigg(\frac{\chi_3^\prime}{2}+\frac{2}{r}\bigg)+\chi_4
h^{\ast^{\prime\prime}}\\\nonumber&+\chi_4\chi_3^\prime
h^{\ast^\prime}+\frac{\chi_4\psi
h^{\ast^{\prime^2}}}{2}+\frac{\chi_4^\prime
h^{\ast^\prime}}{2}+\frac{2\chi_4 h^{\ast^\prime}}{r}\bigg]
-g^\ast\bigg(\chi_3^{\prime\prime}+\frac{\chi_3^{\prime^2}}{2}
+\frac{\chi_3^\prime}{r}\bigg)-g^{\ast^\prime}\bigg(\frac{\chi_3^\prime}{2}
+\frac{1}{r}\bigg)\\\label{38}&-\frac{\chi_4^\prime
h{\ast^\prime}}{2}- \chi_4\bigg(h^{\ast^{\prime\prime}}+\frac{\psi
h^{\ast^{\prime^2}}}{2}+\chi_3^\prime
h^{\ast^\prime}+\frac{h^{\ast^\prime}}{r}\bigg)=0.
\end{align}
By applying this equation alongside the relationship given by
Eq.\eqref{35}, we obtain numerical estimates for the deformation
functions \(h^\ast\) and \(g^\ast\), which are depicted in Figure
\textbf{1}. Incorporating these approximations for $h^\ast$ and
$g^\ast$ into the EGD metric \eqref{32}, we construct an extended
version of the Bardeen black hole. It is crucial to highlight that
in all of the graphs, we have used the parametric values for Rastall
and decoupling as $\zeta=0.1$ (solid), $0.4$ (dashed), and
$\psi=0.02$ (orange), $0.04$ (gray), $0.06$ (cyan), $0.08$ magenta,
$0.1$ (black), respectively. For the magnetic charge parameter, we
have used $e=1$. To ensure that the region is visible to an external
observer, we set $M=1$. We emphasize that the values selected for
the Rastall and decoupling parameters were chosen after a rigorous
test of values in which they were found to induce the desired
behavior in the effective matter variables.

The admissible ranges of the model parameters are subject to both
theoretical and observational constraints. Inspection of the trace
of the Rastall field equations reveals that $\zeta=\frac{1}{4}$
leads to a singular behavior and must be excluded. Furthermore,
ensuring that the theory reduces to GR in the Newtonian limit
eliminates the case $\zeta=\frac{1}{6}$. Accordingly, the Rastall
parameter obeys $\zeta\neq\frac{1}{4}$ and $\zeta\neq\frac{1}{6}$.
Observationally, Li et al. \cite{58aa} employed 118 galaxy-galaxy
strong-lensing systems to constrain the magnitude of the Rastall
parameter to about $10^{-3}$. In theoretical analyses, however,
moderately larger values are often adopted to investigate the
response of the field variables to the non-minimal coupling. In this
work we consider $\zeta=0.1,0.4$, which avoid the pathological
values above and yield regular, physically admissible solutions. The
decoupling parameter $\psi$ controls the coupling strength between
the seed and additional sources and typically lies within the finite
interval $0\leq\psi\leq1$, corresponding to mild geometric
deformations that preserve regularity and asymptotic flatness while
clearly reflecting the influence of the auxiliary source.

\begin{figure}\center
\epsfig{file=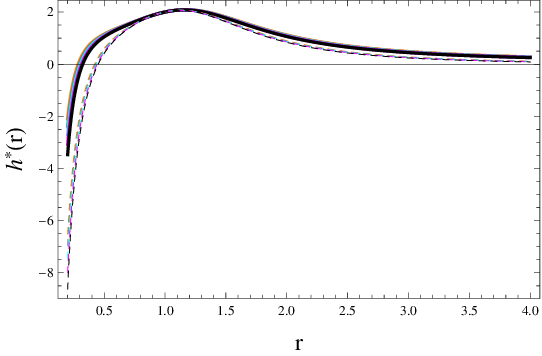,width=0.475\linewidth}
\epsfig{file=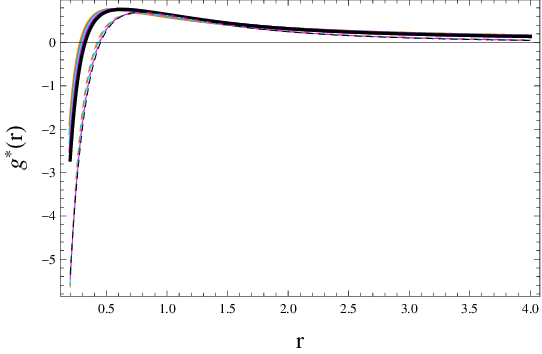,width=0.475\linewidth}\caption{Plots of
$h^\ast(r)$ and $g^\ast(r)$ against $r$ for solution I.}
\end{figure}

An important question is whether the enhanced model maintans the
regularity of the original Bardeen black hole model. By analyzing
the modified metric \eqref{32}, it becomes evident that the
regularity of the enhanced models hinges on the deformation
functions, $h^\ast(r)$ and $g^\ast(r)$. This suggests that the
enhanced model remains regular as long as these deformation
functions are well defined at the core or center. As shown in Figure
\textbf{1}, the deformation functions derived from the initial model
are regular at the center, confirming the regularity of the enhanced
model. Also, we offer a diagram showing the altered metric
coefficients, which facilitates the evaluation of the spacetime
asymptotic flatness. Asymptotic flatness describes the condition
where the metric potentials approach to 1 as the radial distance
becomes infinitely large. In this context, the gravitational field
gradually diminishes and becomes negligible at far distances from a
massive object. Thus, at such vast distances, spacetime approximates
the flat geometry of special relativity, where gravity is absent.
The graph of the modified metric potentials shown in Figure
\textbf{2} indicates that the spacetime is asymptotically flat.
\begin{figure}\center
\epsfig{file=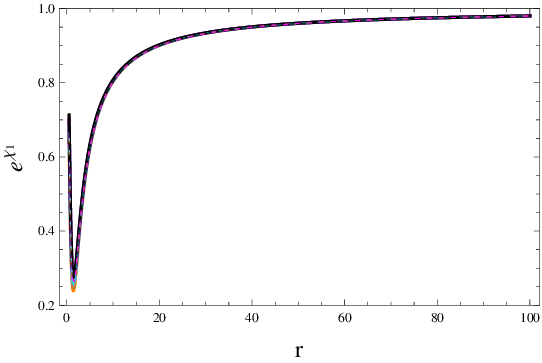,width=0.475\linewidth}
\epsfig{file=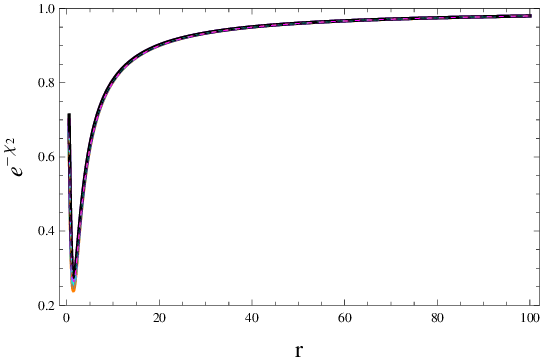,width=0.475\linewidth}\caption{Plots of
$e^\chi_1(r)$ and $e^{-\chi_2(r)}$ against $r$ for solution I.}
\end{figure}
\begin{figure}\center
\epsfig{file=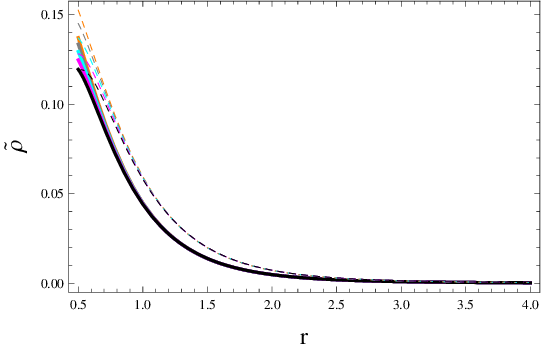,width=0.475\linewidth}
\epsfig{file=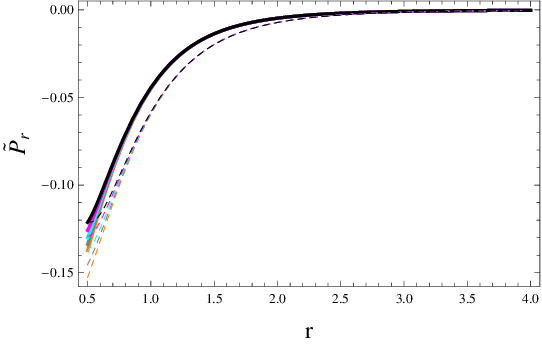,width=0.475\linewidth}
\epsfig{file=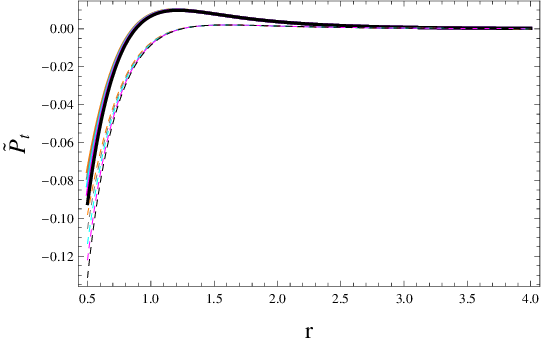,width=0.475\linewidth}
\caption{$\tilde{\rho},\tilde{P}_r$, and $\tilde{P}_t$ for solution
I.}
\end{figure}

The model is presented through its effective parameters in Figure 3.
In this case, we see that the density is above zero and the radial
pressure is below, which are normal for these quantities. Negative
radial pressure indicates an attractive force that enhances the
black hole's gravity. In negative radial pressure theories, these
states are incorporated in order to reason about inflationary and
other aspects, for example about the dark energy which has negative
pressure and is believed to be accelerating the universe. Moreover,
with respect to $\zeta$, the energy density shows a direct
correlation whereas the radial and transverse pressures exhibit an
inverse relationship.  In the case of decoupling parameter $\psi$,
there is slight alteration in the behavior of the parameters
$\tilde{\rho},\tilde{P}_r,$ and $\tilde{P}_t$.
\begin{figure}\center
\epsfig{file=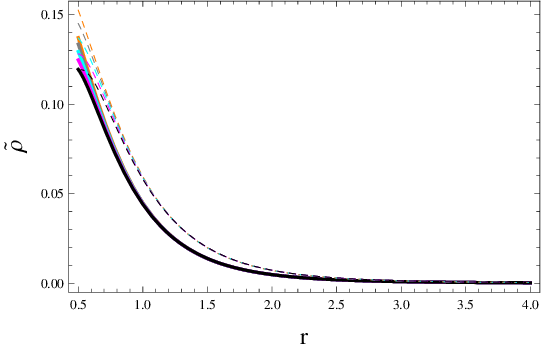,width=0.475\linewidth}
\epsfig{file=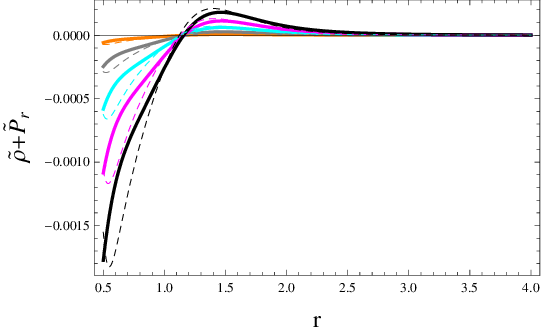,width=0.475\linewidth}
\epsfig{file=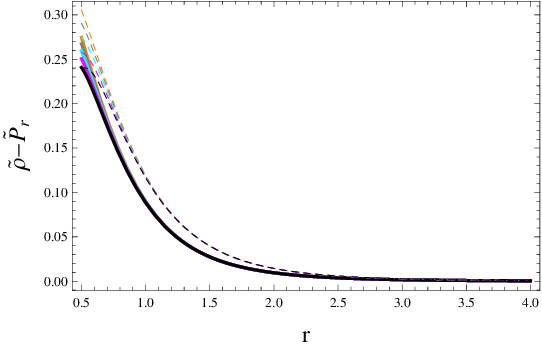,width=0.475\linewidth}
\epsfig{file=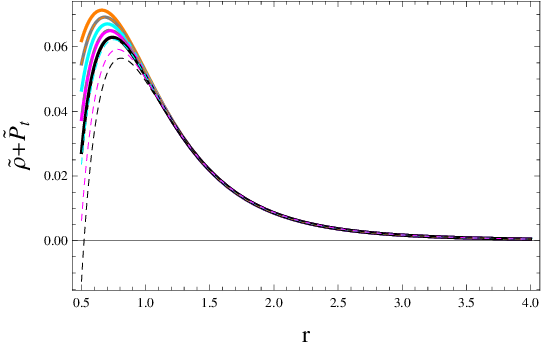,width=0.475\linewidth}
\epsfig{file=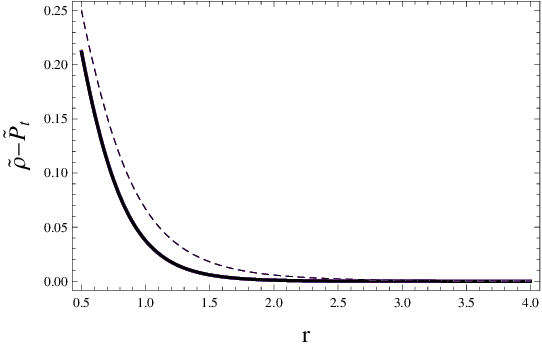,width=0.475\linewidth}
\epsfig{file=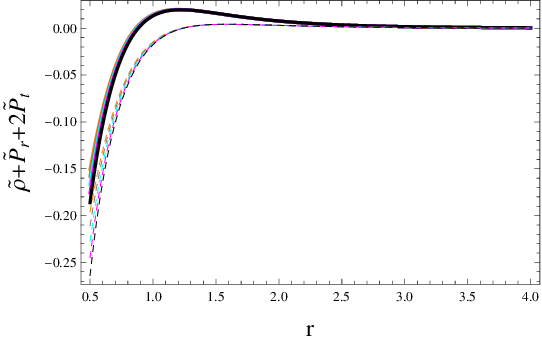,width=0.475\linewidth} \caption{Energy conditions
for solution I.}
\end{figure}

Lastly, we ascertain if the thermodynamic variables satisfy
specified energy bounds given by
\begin{equation}\nonumber
\tilde{\rho}\geq
0,\quad\tilde{\rho}+\tilde{P_r}\geq-2\tilde{P_t},\quad
\tilde{\rho}\geq |\tilde{P_r}|,\quad\tilde{\rho}\geq |\tilde{P_t}|.
\end{equation}
The satisfaction of the above criteria signifies that the matter in
question is conventional. Otherwise, it implies that the matter is
of an unusual or exotic nature. The plots in Figure \textbf{4}
illustrate that the matter source is exotic since some energy
conditions are not met.

\subsection{Solution II: A Barotropic EoS}

The additional source, denoted as $\Omega_{\eta\xi}$, is referred to
as a polytropic fluid if it obeys the EoS given by \cite{58}
\begin{equation}\label{39}
\tilde{P_r} = \lambda \bigg(\tilde{\rho}\bigg)^\beta,
\end{equation}
where $\lambda>0$ encodes parametric information related to the
temperature, and $\beta=1+\frac{1}{n}$, with $n$ representing the
polytropic index. By applying the appropriate substitutions and
considering the special case where $\beta=1$, the equation
simplifies to
\begin{equation}\label{40}
\lambda(\Omega^0_0)+\Omega_1^1=0,
\end{equation}
denoting a barotropic EoS \cite{58}. Equation \eqref{40} can be
identified as a particular case of the EoS \eqref{36}, with
$\mu=\frac{1}{\lambda}$ and $\nu=0$. Using Eqs.\eqref{26} and
\eqref{27}, this equation gives
\begin{figure}\center
\epsfig{file=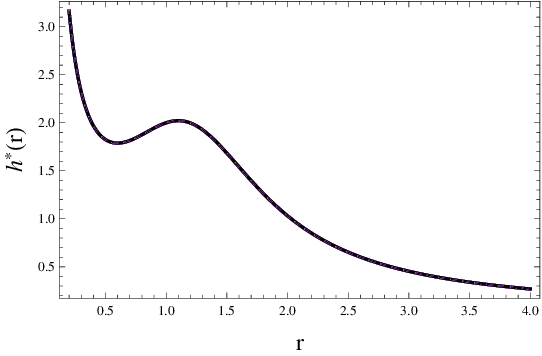,width=0.475\linewidth}
\epsfig{file=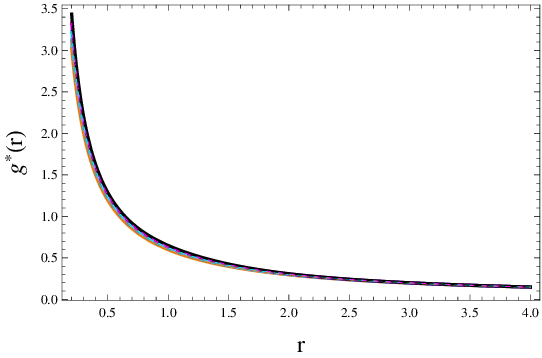,width=0.475\linewidth}\caption{Plots of
$h^\ast(r)$ and $g^\ast(r)$ against $r$ for solution II.}
\end{figure}
\begin{align}\nonumber
&-\lambda\bigg(\frac{g^{\ast^\prime}}{r}+\frac{g^\ast}{r^2}\bigg)
-g^\ast\bigg(\frac{\chi_3^\prime}{r}+\frac{1}{r^2}\bigg)-\frac{\chi_4
h^{\ast^\prime}}{r}+\frac{\zeta(\lambda+1)}{4}\bigg[g^\ast\bigg(\chi_3^{\prime\prime}
+\frac{\chi_3^{\prime^2}}{2}+\frac{2\chi_3^\prime}{r}\bigg)\\\label{41}&
+g^{\ast^\prime}\bigg(\frac{\chi_3^\prime}{2}+\frac{2}{r}\bigg)
+\chi_4 h^{\ast^{\prime\prime}}+\chi_4\chi_3^\prime h^{\ast^\prime}
+\frac{\chi_4\psi h^{\ast^{\prime^2}}}{2}+\frac{\chi_4^\prime
h^{\ast^\prime}}{2}+\frac{2\chi_4 h^{\ast^\prime}}{r}\bigg]=0.
\end{align}
Using this equation along with Eq.\eqref{35}, we obtain numerical
estimates for the deformation functions $h^\ast$ and $g^\ast$. These
functions are then integrated into the EGD metric \eqref{32} to
create a new extended model. This model retains the same effective
matter variables outlined in Eq.\eqref{33}, but now utilizes the
deformation functions derived from Eq.\eqref{41}. As shown in Figure
\textbf{5}, these deformation functions exhibit no singularities
within their domain. Consequently, following the reasons given in
the earlier analysis, it may be deduced that the extended model made
using these deformation functions is regular.
\begin{figure}\center
\epsfig{file=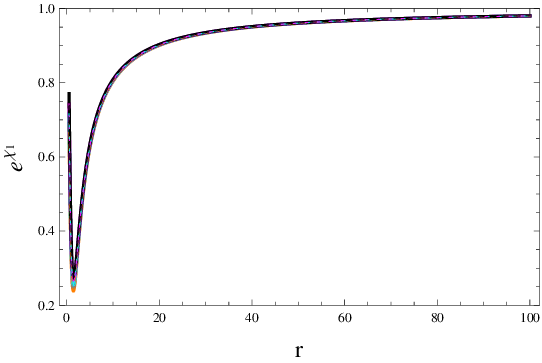,width=0.475\linewidth}
\epsfig{file=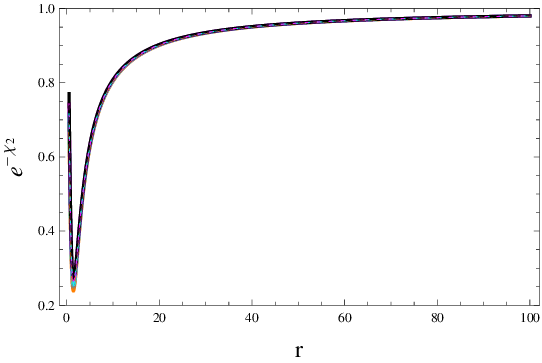,width=0.475\linewidth}\caption{Plots of
$e^\chi_1(r)$ and $e^{-\chi_2(r)}$ against $r$ for solution II.}
\end{figure}
\begin{figure}\center
\epsfig{file=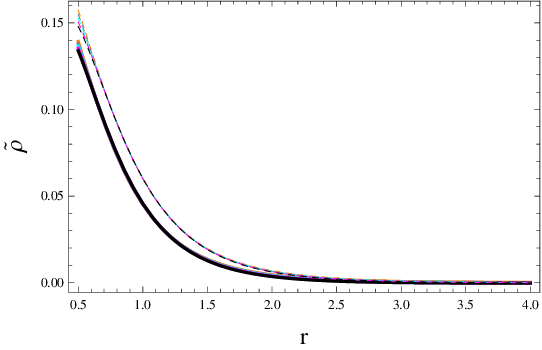,width=0.475\linewidth}
\epsfig{file=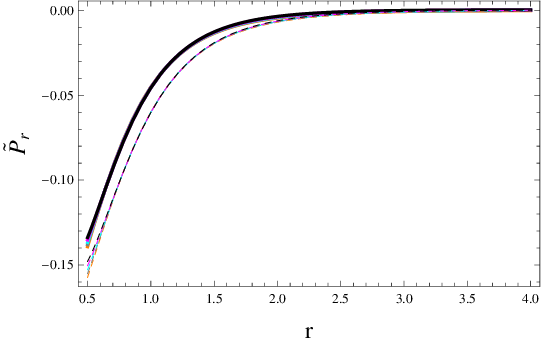,width=0.475\linewidth}
\epsfig{file=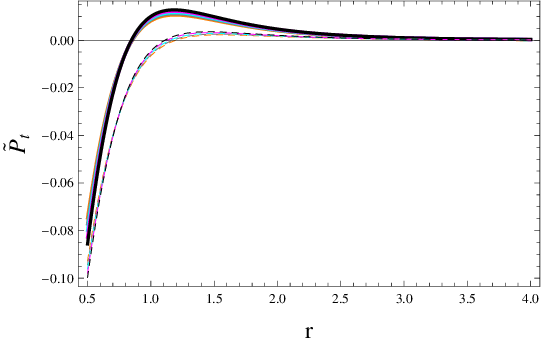,width=0.475\linewidth}\caption{$\tilde{\rho},\tilde{P}_r$,
and $\tilde{P}_t$ for solution II.}
\end{figure}

We also assess the asymptotic flatness of this spacetime using the
same approach as in the previous model. Our analysis confirms that
this spacetime maintains asymptotic flatness (Figure \textbf{6}).
The matter variables shown in Figure \textbf{7} deepen our insights
into the model. We observes a positive $\tilde{\rho}$ and a negative
$\tilde{P}_r$, while $\tilde{P}_t$ fluctuates between negative and
positive values. It is further observed that $\tilde{\rho}$ varies
directly with the Rastall parameter, whereas $\tilde{P}_r$ and
$\tilde{P}_t$ show an inverse relationship. On the other hand, the
effective parameters display minimal variation with respect to
$\psi$. Ultimately, the energy limits are shown in Figure \textbf{8}
where the breach of some conditions indicates the existence of some
exotic source.
\begin{figure}\center
\epsfig{file=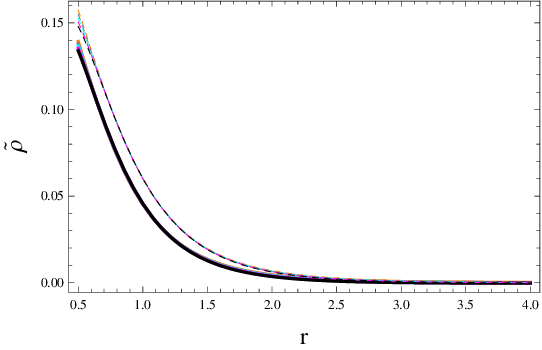,width=0.475\linewidth}
\epsfig{file=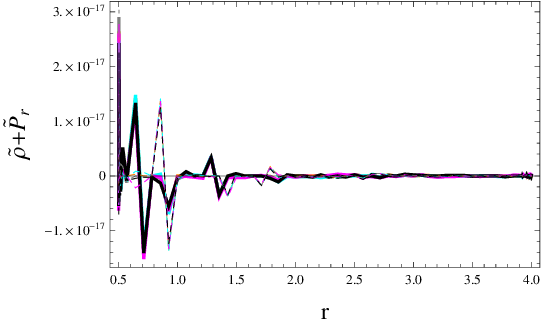,width=0.475\linewidth}
\epsfig{file=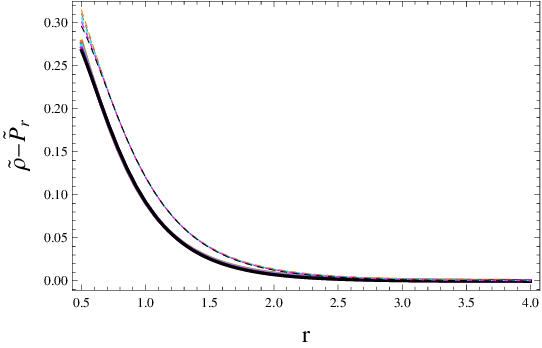,width=0.475\linewidth}
\epsfig{file=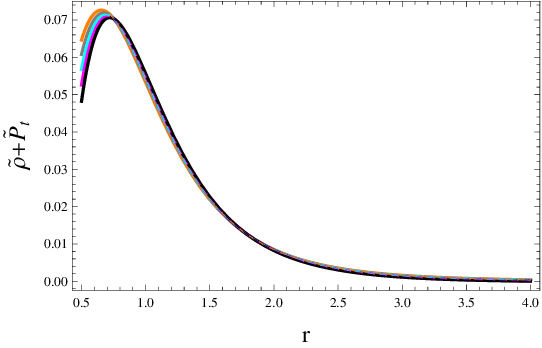,width=0.475\linewidth}
\epsfig{file=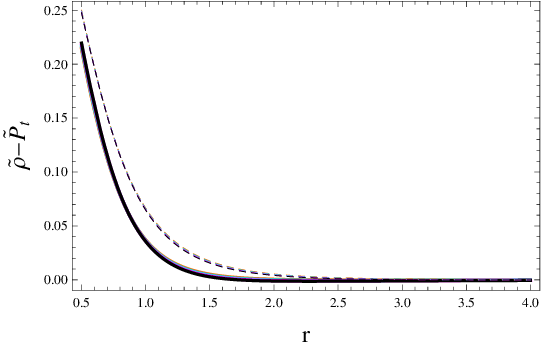,width=0.475\linewidth}
\epsfig{file=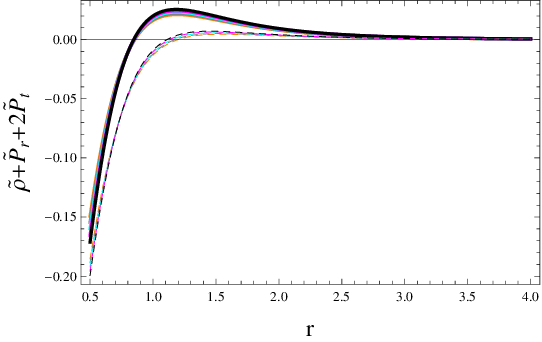,width=0.475\linewidth} \caption{Energy conditions
for solution II.}
\end{figure}

\section{A Thermodynamic Perspective}

Here, we study some thermodynamic properties like temperature,
entropy, and specific heat of black holes, which are examples of
aspects that connect quantum mechanics and GR. Scientists gain a new
understanding of spacetime, quantum field behavior in extreme
gravity, and the laws which dictate the behavior of the universe
based on the process of black holes being heated and cooled due to
the absorption and emission of radiation. This cross-relational
perspective not only sheds more light on how black holes work but
also helps in understanding the extreme laws of physics such as
relativity and thermodynamics.

\subsection{Hawking Temperature}

\begin{equation}\label{42}
T=\frac{1}{4\pi}\bigg |\frac{g_{tt,r}}{\sqrt{-g_{tt}g_{rr}}}\bigg
|_{r=r_H}=\frac{k}{2\pi}.
\end{equation}
\begin{figure}\center
\epsfig{file=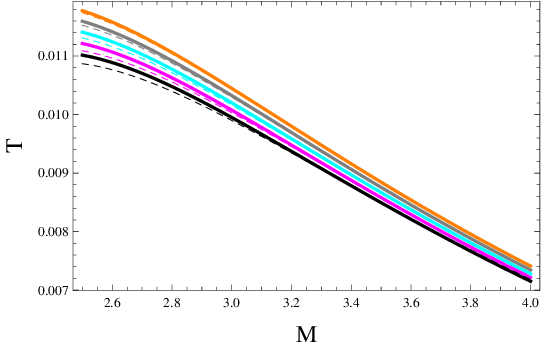,width=0.5\linewidth}\caption{Plot of $T$ against
$M$ for solution I.}
\end{figure}

In Figure \textbf{9}, we illustrate the temperature curve for
solution I, which demonstrates the anticipated behavior by showing
an inverse correlation between the mass and temperature of the black
hole. Close to the center, the Rastall parameter, $\zeta$ shows a
direct variation to the temperature. On the other hand, the
decoupling parameter is strained in inverse proportion to the
temperature of radiation. Likewise, the Hawking temperature of
solution II which is shown in Figure \textbf{10} shows a correct
behavior. Contrary to the first model, the fluctuation in the
Rastall parameter does not make a difference. Still, as in the
earlier model, the decoupling parameter is once more inversely
proportional to the temperature.
\begin{figure}\center
\epsfig{file=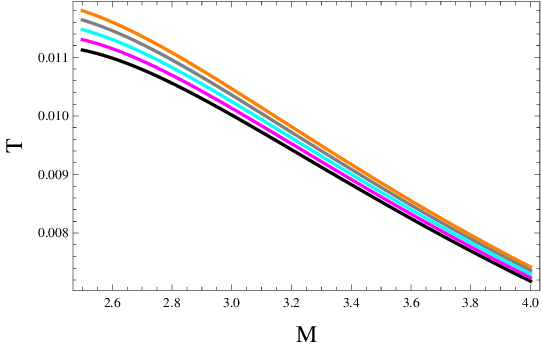,width=0.5\linewidth}\caption{Plot of $T$ against
$M$ for solution II.}
\end{figure}

\subsection{Specific Heat}

An important thermodynamic variable that is employed in the analysis
of the thermal stability of black holes, is the specific heat
parameter. The specific heat represents the amount of heat that is
required to change the temperature of a black hole by a small
amount. In black hole thermodynamics, it exists as a measure of
stability, a black hole that has a positive specific heat is one
that can be brought into thermal contact with other black holes
without drying out, hence it is stable. On the other hand, a black
hole with a negative specific heat will not be able to sustain
heating or cooling during heat transfer, hence leads to dissipation
of heat which depicts an unstable situation. This is particularly
clear in the case of different types of black holes such as the
Schwarzschild and Kerr black holes. This quantity is expressed as
\begin{equation}\label{43}
C=T\bigg(\frac{\partial S}{\partial T}\bigg)\bigg
|_{r=r_H}=T\bigg(\frac{\partial S}{\partial
r_H}\bigg)\bigg(\frac{\partial T}{\partial r_H}\bigg)^{-1},
\end{equation}
where
\begin{equation}\label{44}
S=\frac{1}{4}\int_0^{2\pi}\int_0^\pi\sqrt{g_{\theta\theta}
g_{\phi\phi}}d\theta d\phi=\pi r_H^2,
\end{equation}
\begin{figure}\center
\epsfig{file=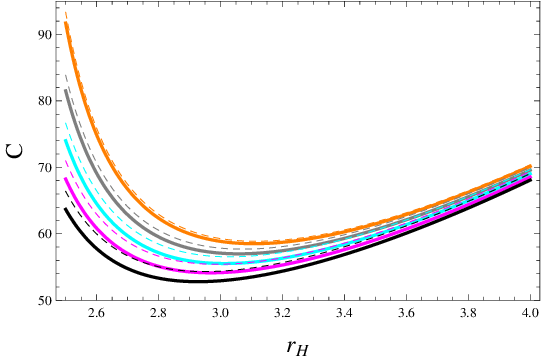,width=0.5\linewidth}\caption{Plot of $C$ against
$r_H$ for solution I.}
\end{figure}
is the entropy. In modified gravity theories like Rastall's, where
the energy-momentum tensor is not conserved, the entropy may acquire
corrections beyond the standard Bekenstein-Hawking area law.
However, the formulation of a Euclidean action for Rastall gravity
is non-trivial due to the absence of a canonical variational
principle. While we adopt the standard entropy expression here, we
acknowledge the importance of a rigorous derivation. In the absence
of an explicit action, one possible direction for future work
involves employing thermodynamic potentials derived from quasi-local
energy definitions, such as the Misner-Sharp mass, or analyzing
horizon thermodynamics directly under the non-conservation
condition. These approaches could potentially reveal entropy
corrections induced by the Rastall parameter, and we intend to
investigate them in subsequent studies.

For the first model Figure \textbf{11} demonstrates that the
specific heat is positive within the range $2.5 \leq r_H \leq 4$,
indicating stability in this range. It is evident that $\zeta$
exhibits a direct relationship with specific heat, while $\psi$
shows an inverse relationship. In the case of the second model
(Figure \textbf{12}), there is an observed inverse relationship
between $\psi$ and the specific heat, while $\zeta$ remains
unaffected by the specific heat. This model also indicates stability
within the range $2.5 \leq r_H \leq 4$.
\begin{figure}\center
\epsfig{file=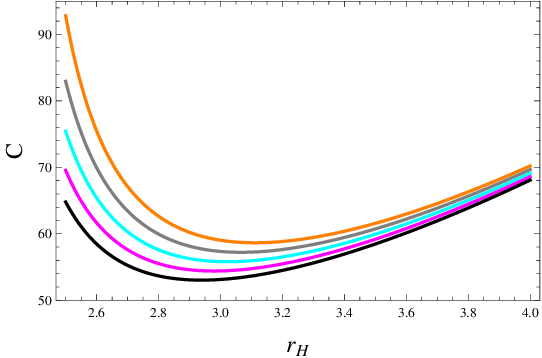,width=0.5\linewidth}\caption{Plot of $C$ against
$r_H$ for solution II.}
\end{figure}

\subsection{Hessian Matrix}

The Hessian matrix is particularly important to the study of the
thermodynamic stability of black holes when looking at its trace. It
is comprised of partial derivatives of Helmholtz free energy,
$F=E-ST$, in which $E$, $S$ and $T$ refer to the internal energy,
entropy and the temperature of the black hole, respectively.
Derivatives of these are derived concerning temperature and volume.
For the purposes of this exercise the temperature employed is the
Hawking temperature and the volume of the black hole is defined as
$V=\frac{4}{3}\pi r_H^3$. The Hessian matrix is expressed as
\begin{equation}\label{45}
H=\begin{pmatrix}
h_{11} & h_{12} \\
\\
h_{21} & h_{22}
\end{pmatrix}
=
\begin{pmatrix}
\frac{\partial^2F}{\partial T^2} & \frac{\partial^2F}{\partial T\partial V} \\
\\
\frac{\partial^2F}{\partial V\partial T} &
\frac{\partial^2F}{\partial V^2}
\end{pmatrix}.
\end{equation}
We explore the trace
\begin{equation}\label{46}
Tr(H)=h_{11}+h_{22},
\end{equation}
to determine the stability of our models, with $Tr(H)\geq 0$
denoting stability \cite{59}.
\begin{figure}\center
\epsfig{file=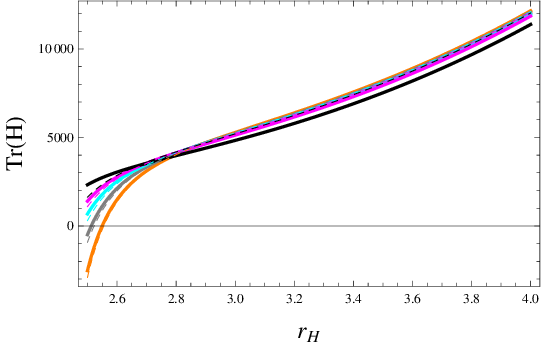,width=0.5\linewidth}\caption{Plot of $Tr(H)$ against
$r_H$ for solution I.}
\end{figure}
\begin{figure}\center
\epsfig{file=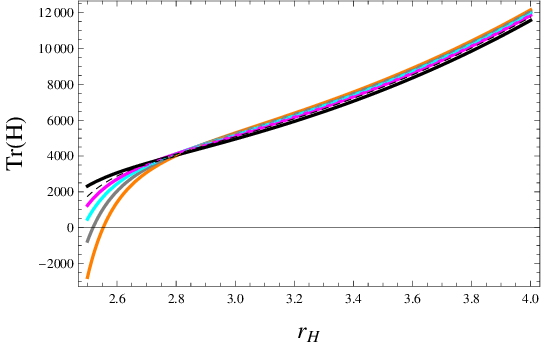,width=0.5\linewidth}\caption{Plot of $Tr(H)$ against
$r_H$ for solution II.}
\end{figure}

The graphical representation of $Tr(H)$ for the initial model is
illustrated in Figure \textbf{13}, highlighting the model's
stability within the range $2.55 \leq r_H \leq 4$. Interestingly,
the fluctuations in the Rastall parameter are either absent or
minimal, while the decoupling parameter shows a divergent trend with
$Tr(H)$. For the second model, depicted in Figure \textbf{14},
stability is also noted within the interval $2.55 < r_H \leq 4$.
Moreover, the behavior of the Rastall and decoupling parameters
relative to the trace mirrors that of the previous model.

\section{Conclusions}

The study in this paper is geared towards obtaining generalizations
of a well-known regular black hole solution, namely the Bardeen
black hole. To this end, we utilize the well known gravitational
decoupling scheme which is a powerful tool used in obtaining new and
extended solutions from known solutions. With this scheme, extended
solutions that preserve fundamental properties of the parent ansatz
can be obtained. The field equations with an extra matter source,
$\Omega_{\eta\xi}$, are extensively decoupled by employing geometric
deformations via a linear transformation. The source
$\Omega_{\eta\xi}$ facilitates the extension of the Bardeen black
hole, yielding new regular black hole models. The decoupling process
leads to two new subsystems, the first of which is specified by the
Bardeen black hole while the second set characterized by
$\Omega_{\eta\xi}$ is resolved via a generalized constraint given by
$\Omega_0^0+\mu\Omega_1^1=\nu\Omega^2_2,~\mu,\nu\in\mathbb{R}$.

We obtain two novel black hole models which are analyzed in detail
for different physical properties. We also study in detail, the
impact of the decoupling and Rastall parameters on the obtained
models. The matter variables, particularly the density and radial
pressure were found to execute viable behavior in both models.
Additionally, both models were found to preserve asymptotic
flatness, while a violation of some energy bounds were observed. Due
to this violation, our models suggest black holes characterized by
exotic matter. The analysis of the radiation temperature revealed
that for black holes of lower mass, the level of radiation emitted
was greater. This result is in line with the expectations since the
emission of radiation leads to the evaporation and hence decrease in
mass. The Rastall parameter presents a slight reverse dependence to
the temperature in the first model. This dependence can be noticed
around the core region. In the second model however, the Rastall
parameter presents no dependence to the temperature. In both models,
the decoupling parameter is inversely related to the temperature.
Lastly, we have analyzed the thermodynamic stability in terms of the
heat capacity and trace, $Tr(H)$. The results of these test
considered together suggest that both models are stable for all
values within the range $2.55\leq r_H\leq4$.

A crucial distinction between our current study and our previous
work on minimally deformed Bardeen black holes in Rastall gravity
\cite{58ab} is the preservation of asymptotic flatness. In the
previous study, both models failed to maintain asymptotic flatness,
while in this work, the extended gravitational decoupling approach
ensures that both solutions remain asymptotically flat. This
contrast highlights the impact of different decoupling methods on
the large-scale properties of black hole solutions in modified
gravity. These findings further solidify the novelty of this study
and its contributions to black hole physics in non-conservative
gravity frameworks.\\\\
\textbf{Data Availability Statement:} No new data were generated or
analyzed in support of this research.


\begin{thebibliography}{00}

\bibitem{5} Rastall, P.: Phys. Rev. D \textbf{6}(1972)3357.

\bibitem{6} Parker, L.: Phys. Rev. D \textbf{3}(1971)346.

\bibitem{7} Gibbons, G.W. and Hawking, S.W.: Phys. Rev. D \textbf{15}(1977)2738.

\bibitem{8} Ford, L.H.: Phys. Rev. D \textbf{35}(1987)2955.

\bibitem{1a} Visser, M.: Phys. Lett. B \textbf{782}(2018)83.

\bibitem{1aa} Golovnev, A.: Ann. Phys. \textbf{461}(2024)169580.

\bibitem{1b} Darabi, F. et al.: Eur. Phys. J. C \textbf{78}(2018)25;
Darabi, F., Atazadeh, K. and Heydarzade, Y.: Eur. Phys. J. Plus
\textbf{133}(2018)249; Hansraj, S., Banerjee, A. and Channuie, P.:
Ann. Phys. \textbf{400}(2019)320.

\bibitem{2} Singh, A. and Mishra, K.C.: Eur. Phys. J. Plus
\textbf{135}(2020)752.

\bibitem{3} Singh, A., Raushan, R. and Chaubey, R.: Can. J. Phys.
\textbf{99}(2021)1073.

\bibitem{4} Singh, A., Singh, G.P. and Pradhan, A.: Int. J. Mod. Phys. A
\textbf{37}(2022)2250104; Singh, A. and Pradhan, A.: Indian J. Phys.
\textbf{97}(2023)631.

\bibitem{4a} Sharif, M. and Sallah, M.: Mod. Phys. Lett. A
\textbf{40}(2025)2550207.

\bibitem{5a} Naseer, T.: Phys. Dark Universe \textbf{46}(2024)101663.

\bibitem{5aa} Naseer, T.: Eur. Phys. J. C \textbf{84}(2024)01.

\bibitem{5aaa} Naseer, T. and Sharif, M.: Class. Quantum Grav. \textbf{41}(2024)245006.

\bibitem{14} Abbott, B.P. et al.: Phys. Rev. Lett. \textbf{116}(2016)061102.

\bibitem{15} Abbott, B.P. et al.: Phys. Rev. Lett. \textbf{119}(2017)141101.

\bibitem{16} Akiyama, K. et al.: Astrophys. J. Lett. \textbf{875}(2019)L2.

\bibitem{17} Hawking, S.W.: Commun. Math. Phys. \textbf{25}(1972)152.

\bibitem{18} Ruffini, R. and Wheeler, J.A.: Phys. Today \textbf{24}(1971)30.

\bibitem{18a} Hawking, S.W., Perry, M.J. and Strominger, A.: Phys. Rev. Lett.
\textbf{116}(2016)231301.

\bibitem{19} Penrose, R.: Phys. Rev. Lett. \textbf{14}(1965)57.

\bibitem{20} Hawking, S.W.: Phys. Rev. Lett. \textbf{15}(1965)689.

\bibitem{21} Hawking, S.W.: Proc. Roy. Soc. Lond. A \textbf{294}(1966)511.

\bibitem{22} Hawking, S.W.: Proc. Roy. Soc. Lond. A \textbf{295}(1966)490.

\bibitem{23} Hawking, S.W.: Proc. Roy. Soc. Lond. A \textbf{300}(1967)187.

\bibitem{24} Israel, W.: Commun. Math. Phys. \textbf{8}(1968)245.

\bibitem{25} Hawking, S.W.: Nature \textbf{248}(1974)30.

\bibitem{26} Hawking, S.W.: Commun. Math. Phys. \textbf{43}(1975)199.

\bibitem{27} Bardeen, J.: Proc. GR5, (Tiflis, USSR, 1968).

\bibitem{28} Ayon-Beato, E. and Garcia, A.: Phys. Rev. Lett. \textbf{80}(1998)5056.

\bibitem{29} Hayward, S.A.: Phys. Rev. Lett. \textbf{96}(2006)031103.

\bibitem{30} Zaslavskii, O.B.: Phys. Lett. B. \textbf{688}(2010)278.

\bibitem{31} Narzilloev, B. et al.: Phys. Rev. D \textbf{102}(2020)104062.

\bibitem{32} Culetu, H.: Int. J. Mod. Phys. D \textbf{31}(2022)2250124.

\bibitem{33} Balart, L., Panotopoulos, G. and Rinc{\'o}n, A.: Fortschr. Phys. \textbf{71}(2023)2300075.

\bibitem{34} Ovalle, J., Casadio, R. and Giusti, A.: Phys. Lett. B \textbf{844}(2023)138085.

\bibitem{35} Bekenstein, J.D.: Phys. Rev. D \textbf{9}(1974)3292.

\bibitem{36} Heydarzade, Y., Moradpour, H. and Darabi, F.: Can. J. Phys. \textbf{95}(2017)1253.

\bibitem{37} Kumar, R. and Ghosh, S.G.: Eur. Phys. J. C \textbf{78}(2018)750.

\bibitem{38} Lobo, I.P. et al.: Int. J. Mod. Phys. D \textbf{27}(2018)1850069.

\bibitem{39} Prihadi, H.L. et al.: Int. J. Mod. Phys. D \textbf{29}(2020)2050021.

\bibitem{40} Toshmatov, B., Stuchl{\'i}k, Z. and Ahmedov, B.: Phys. Dark Universe \textbf{41}(2023)101257.

\bibitem{41} Ovalle, J.: Mod. Phys. Lett. A \textbf{23}(2008)3247.

\bibitem{42} Morales, E. and Tello-Ortiz, F.: Eur. Phys. J. C \textbf{78}(2018)841.

\bibitem{43} Tello-Ortiz, F. et al.: Eur. Phys. J. C \textbf{79}(2019)885.

\bibitem{44} da Rocha, R.: Eur. Phys. J. C \textbf{81}(2021)845; ibid. \textbf{82}(2022)34.

\bibitem{45} Maurya, S.K. et al.: Eur. Phys. J. C \textbf{83}(2023)317.

\bibitem{46} Rehman, H. and Abbas, G.: Chin. Phys. C \textbf{47}(2023)125106.

\bibitem{47} Sharif, M. and Waseem, A.: Chin. J.
Phys. \textbf{60}(2019)426; Ann. Phys. \textbf{405}(2019)14; Sharif,
M. and Majid, A.: Chin. J. Phys. \textbf{68}(2020)406; Phys. Dark
Universe \textbf{30}(2020)100610; Sharif, M. and Saba, S.: Chin. J.
Phys. \textbf{63}(2020)348; Int. J. Mod. Phys. D
\textbf{29}(2020)2050041; Sharif, M. and Sallah, M.: New Astron.
\textbf{109}(2024)102198.

\bibitem{47a} Maurya S.K. and Tello-Ortiz, F.: Phys. Dark Universe
\textbf{29}(2020)100577.

\bibitem{48} Ovalle, J.: Phys. Lett. B \textbf{788}(2019)213.

\bibitem{49} Contreras, E. and Bargue{\~n}o, P.: Class. Quantum Grav.
\textbf{36}(2019)215009.

\bibitem{50} Sharif, M. and Ama-Tul-Mughani, Q.: Ann. Phys. \textbf{415}(2020)168122.

\bibitem{51} Sharif, M. and Majid, A.: Phys. Dark Universe \textbf{30}(2020)100610.

\bibitem{52} Ovalle, J. et al.: Phys. Dark Universe \textbf{31}(2021)100744.

\bibitem{53} Sharif, M. and Majid, A.: Phys. Dark Universe \textbf{32}(2021)100803.

\bibitem{54} Sharif, M. and Majid, A.: Phys. Scr. \textbf{96}(2021)035002.

\bibitem{55} Sharif, M. and Naseer, T.: Indian J. Phys. \textbf{96}(2022)4373;
Int. J. Mod. Phys. D \textbf{31}(2022)2240017; Class. Quantum Grav.
\textbf{40}(2023)035009; Chin. J. Phys. \textbf{86}(2023)596; Eur.
Phys. J. Plus \textbf{139}(2024)86.

\bibitem{56} Sharif, M. and Sallah, M.: Astron. Comput. \textbf{50}(2025)100897.

\bibitem{56aa} Sharif, M. and Sallah, M.: High Energy Density Phys. \textbf{56}(2025)101208.

\bibitem{56d} Cai, R.G., Cao, L.M. and Pang, D.W.: Phys. Rev. D \textbf{72}(2005)044009.

\bibitem{56dd} Ali, M.S., Ghosh, S.G. and Wang, A.: Phys. Rev. D \textbf{108}(2023)044045.

\bibitem{56ddd} Pourhassan, B., Eslamzadeh, S., Sakalli, I. and Upadhyay, S.: arXiv preprint
arXiv:2407.08768.

\bibitem{56e} Soroushfar, S., Pourhassan, B. and Sakalli, I.: Phys. Dark Universe
\textbf{44}(2024)101457.

\bibitem{56ee} Hazarika, B. and Phukon, P.: Nucl. Phys. B \textbf{1012}(2025)116837.

\bibitem{56a} Maurya, S.K. and Tello-Ortiz, F.: Phys. Dark Universe \textbf{29}(2020)100577.

\bibitem{56b} Naseer, T. and Sharif, M.: Universe \textbf{8}(2022)62.

\bibitem{56c} Hassan, K. and Sharif, M.: Universe \textbf{9}(2023)165.

\bibitem{58} Ovalle, J. et al.: Eur. Phys. J. C \textbf{78}(2018)960.

\bibitem{58a} Sharif, M. and Sallah, M.: Phys. Scr. \textbf{99}(2024)115031

\bibitem{58ab} Sharif, M. and Sallah, M.: Chin. J. Phys. \textbf{92}(2024)794.

\bibitem{58aa} Li, R.; Wang, J.; Xu, Z.; Guo, X.: Mon. Not. R. Astron. Soc.
\textbf{486}(2019)2407.

\bibitem{59} Cuadros-Melgar, B. et al.: Eur. Phys. J. C \textbf{80}(2020)848.

\bibitem{60} Misyura, M., Rincon, A. and Vertogradov, V.:
Phys. Dark Universe \textbf{46}(2024)101717.

\end{thebibliography}
\end{document}